# On SDN-Enabled Online and Dynamic Bandwidth Allocation for Stream Analytics

Walid Aljoby, *Student Member, IEEE*, Xin Wang, Tom Z. J. Fu, and Richard T. B. Ma, *Senior Member, IEEE*

*Abstract*—Data communication in cloud-based distributed stream data analytics often involves a collection of parallel and pipelined TCP flows. As the standard TCP congestion control mechanism and its variants are designed for achieving "fairness" among competing flows and are agnostic to the application layer contexts, the bandwidth allocation among a set of TCP flows traversing bottleneck links often leads to sub-optimal application-layer performance measures, e.g., stream processing throughput or average tuple complete latency. Motivated by this and enabled by the rapid development of the software-defined networking (SDN) techniques, in this paper, we re-investigate the design space of the bandwidth allocation problem and propose a cross-layer framework which utilizes the instantaneous information obtained from the application layer and provides on-the-fly and dynamic bandwidth adjustment algorithms for assisting the stream analytics applications achieving better performance during the runtime. We implement a prototype cross-layer bandwidth allocation framework based on a popular open-source distributed stream processing platform, Apache Storm, together with the OpenDaylight controller, and carry out extensive experiments with real-world analytical workloads on top of a local cluster consisting of ten workstations interconnected by a SDN-enabled fat-tree like testbed. The experiment results clearly validate the effectiveness and efficiency of our proposed framework and algorithms. Finally, we leverage the proposed cross-layer SDN framework and introduce an exemplary mechanism for bandwidth sharing and performance reasoning among multiple active applications and show a case of a point solution on how to approximate application-level fairness.

Manuscript received October 11, 2018; revised June 27, 2019; accepted June 29, 2019. Date of publication July 5, 2019; date of current version August 6, 2019. This work was supported in part by Singapore MoE under Grant R-252-000-A67-114, by the National Research Foundation, Prime Ministers Office, Singapore under its Corporate Laboratory @ University Scheme, National University of Singapore, and Singapore Telecommunications Ltd. and in part by the research grant from Singapore Agency for Science, Technology and Research (A*STAR) through the SINGA Program. T. Fu was supported in part by the Natural Science Foundation of China under Grant 61702113, and in part by the China Postdoctoral Science Foundation under Grant 2017M612613. A preliminary report of this work was presented at the IEEE ICNP 2018 Conference [1]. *(Corresponding author: Richard T. B. Ma.)*

W. Aljoby is with the Department of Computer Science, National University of Singapore, Singapore 119077, and also with the Advanced Digital Sciences Center, University of Illinois Research Center, Singapore 138602 (e-mail: algobi@comp.nus.edu.sg).

X. Wang and R. T. B. Ma are with the Department of Computer Science, National University of Singapore, Singapore 119077 (e-mail: dcswan@nus.edu.sg; tbma@comp.nus.edu.sg).

T. Z. J. Fu is with the Advanced Digital Sciences Center, University of Illinois Research Center, Singapore 138602 (e-mail: tom.fu@adsc-create.edu.sg).

Color versions of one or more of the figures in this paper are available online at http://ieeexplore.ieee.org.

Digital Object Identifier 10.1109/JSAC.2019.2927062

## I. INTRODUCTION

LARGE-SCALE stream processing has recently gained high importance due to a large variety of supported applications such as business intelligence, video analytics, machine learning, and event monitoring and detection. These applications process high volumes of unbounded with unpredictable and variable data streams. Nevertheless, they entail a variety of processing requirements, and thus researchers and practitioners have been developing a diverse array of stream processing frameworks (e.g., Storm [2], Heron [3], Samza [4], Flink [5], MillWheel [6], and Ares [7]) to meet the increasing demands of these applications.

For streaming applications, the essential performance indicator of how the application reacts to the incoming data streams is the time needed for each of them to be completely processed. To help the application to achieve a desirable performance characteristics (e.g., delivering real-time response), stream processing frameworks need to effectively and dynamically allocate system resources including CPU, memory, and bandwidth among application components (instances and their flows) in order to expose a highly-optimized pipeline (i.e., execution path) [8], [9].

Yet, many applications today are data-intensive, as opposed to compute-intensive [10]. Indeed, in data-intensive applications, stream processing involves a higher network resource demands than CPU cycles, particularly when the data stream ingestion rates or derived tuples rates from these streams are higher than provisioned network bandwidth. As such, transfer across the network might be the cause of performance bottleneck rather than CPU cycles, therefore managing and optimizing network activity is important to improving and delivering real-time responses in these applications. In this context, there has been flurry of research attempts toward optimizing streaming applications. While in large part successful, however, their focus mainly has centered to schedule and provision computation resources of the applications or limited to minimizing traffic across the network. Hence, these solutions have largely overlooked allocation and provision of network bandwidth. As a result, they are either suboptimal in optimizing network transfer [9], [11], [12], or assuming the network with sufficient bandwidth resource [13].

In current stream processing frameworks, the share of network bandwidth has left to the mercy of the underlying transport mechanisms (e.g., TCP [14], DCTCP [15]). Nonetheless, such mechanisms are designed mainly for end-to-end data delivery in an application agnostic manner, i.e., flows traversing the bottleneck links sharing equal portion of the bandwidth. This, with high probability, will lead to sub-optimality in the overall application-level performance because some flows can be of paramount importance than other flows of the same application.

In this paper, we explore the design space of the bandwidth allocation, formulate it as a utility maximization problem, and propose a heuristic algorithm to derive the close-to-optimal solutions. This whole procedure is encapsulated into a cross-layer framework which utilizes the additional information measured from the running applications and quickly deploys the new allocation decisions to the physical network layer. The latter is enabled by the rapid development of the Software-Defined Networking (SDN) techniques and toolkits and realized through plugging in a control plane module developed (in ODL controller [16]) by us. The main contributions we have made in this paper are listed as follows:

1. We formulate the bandwidth allocation among flows belonging to a stream processing application as an optimization problem and design a heuristic algorithm to seeking for the optimal allocation solution.
2. Leveraging the SDN capabilities, we develop a native SDN control plane application that deploys and updates the bandwidth allocation results derived by our optimization algorithm.
3. We develop an automated cross-layer bandwidth allocation framework and implement a prototype of it based on a popular open-source stream processing platform, Apache Storm, integrating with the OpenDaylight SDN controller.
4. We carry out comprehensive performance evaluation through running stream data applications with real-world workloads in a local cluster composed of 10 workstations interconnected by a hardware SDN-enabled switch.
5. We introduce an exemplary mechanism for bandwidth sharing and reasoning of performance among multiple active applications and present a case on how to approximate application-level fairness.
6. We built a fat-tree like testbed to carry out the evaluation of optimization particulars in a more general setting with a multi-hop network.

The rest of the paper is organized as follows. Section II presents briefly an overview of stream processing, model of datacenter fabric, communication flow, and an example stream application motivating our contribution. In Section III, we introduce a formulation of bandwidth allocation as an application-oriented utility maximization problem and highlight surrounding challenges. We then present the details of solution model and optimization framework in Section IV. Section V describes briefly the end-to-end cross-layer SDN-based implementation of proposed solution. Experimental results are presented in Section VI. Also, a preliminary investigation of the performance of multiple applications and the initial results are presented in Section VII. Finally, we explore related works in Section VIII and conclude the paper in Section IX.

## II. BACKGROUND AND MOTIVATION

### A. Stream Processing

*1) Distributed Stream Processing Frameworks:* Main examples of them such as Storm [2], Twitter [3], Samza [4], Flink [5] and MillWheel [6], have been widely adopted in cloud-based data analytics to enable stream processing in a distributed manner with low latency. Using these frameworks, a variety of stream applications are developed for processing continuously arrived data streams from external producers (e.g., web logs, software logs, scalar sensors, and video cameras) through a pipeline of processing stages. Towards this, multiple models such as one-at-a-time and micro-batched have been proposed to cope with diverse stream application requirements [17]. In this paper, however, our focus is particularly pointed towards the one-at-a-time model which accomplishes processing on an individual tuple basis, for delay-sensitive and data-intensive stream applications.

*2) Application Model:* Our conceptual viewpoint in designing our bandwidth allocation mechanism is by abstracting out the entire application as a sequence of Fork-Join stages. In data analytics literature, the application (a.k.a., job) is typically characterized by a logical topology, that defines a dataflow programming paradigm in a form of directed acyclic graph (DAG) of operators (i.e., vertices), through which data streams (i.e., edges) are constantly produced and consumed. Despite differences in DAG structure, we observe that stream applications have one thing in common: Fork-Join pattern, in which each edge starts with a fork operator and end with a join operator. Additionally, each operator can typically be classified according to number of input and output streams into a) *1:1* operator, b) *m:1* operator, and c) *1:m* operator. This variety in operators enables application programmer to flexibly chain them according to the logic of stream application. In Figure 1a, we show an example topology of finding trending tags at LinkedIn [4], which consists of six operators. In this topology, *Split* is a *1:m* type operator constitutes the entry point of the application, typically called source operator. It consumes streams of user profile updates, splits them into skill and job updates and then emits them to the two downstream operators named *Skill Extractor* and *Job Extractor*. Both *Skill Extractor* and *Job Extractor* are of the type *1:1* operators performing tag extraction on the input streams and sending the processing results to the *Merge* operator. The latter is a *m:1* operator combining skill and job tag streams into one and sending them to a *1:1 Count* operator which maintains the frequency of each distinct tag. Lastly, *TopK*, as the last operator in the topology, typically called sink operator, partitions received tag counts into windows (e.g., find top $k$ tags over a 5-second window) and keeps updating application statistics (e.g., trending tags such as user skills and job positions, in this example).

Furthermore, stream applications exhibit a wide diversity in terms of scale, state, and lifetime. In particular, the scaling of stream application is important to cope with computational need of time-constrained applications. As such, processing frameworks offer users a set of APIs to configure multiple instances per operator in order to execute user-defined logic concurrently. Figure 1b presents a parallelization of DAG shown in Figure 1a where each operator is replicated twice (e.g., $Split_1$ and $Split_2$ are instances of the *Split* operator), except the sink operator has only one instance, $TopK_1$.

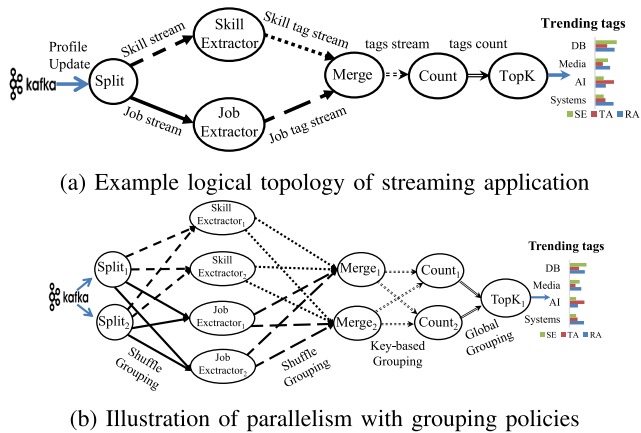

Fig. 1. A Streaming application example (a) Operators component of trending tags logical topology (b) Parallelism of logical topology into instances and data grouping policies among them.

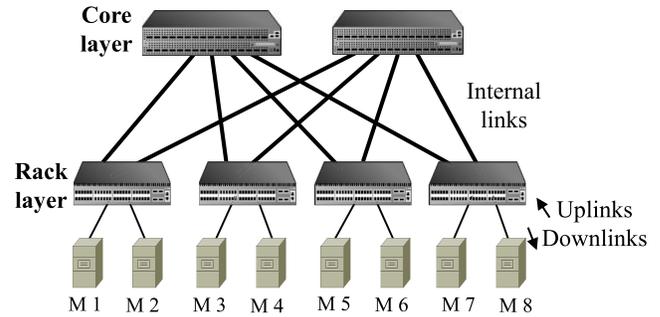

Fig. 2. A datacenter network example with 8 uplinks, 8 downlinks, 16 internal links (8 rack-to-core links, and 8 core-to-rack links).

*3) Stream Grouping Policies:* As parallelizing operators is a key factor to speeding up stream processing, the transfer of data streams between the instances must follow appropriate grouping policies, sometimes called data routing, to meet different application requirements. The main grouping policies of regulating how data tuples are forwarded to target instance set of the downstream operators are categorized as follows.

*a) Shuffle Grouping:* data tuples are sent in a round-robin fashion to the instances of the downstream operator. This policy ensures that the processing workloads are evenly distributed among these receiving instances.

*b) Key-Based Grouping:* the destination instances to which data tuples are sent is determined by applying some well defined projection function, e.g., hashing, on the key (or signature) of the data tuple. This grouping policy has the property that a) the tuples sharing the same key are guaranteed to send to the same instance, i.e., avoiding an additional step for key-based result aggregation; and b) roughly even-partition the key space among all the destination instances, but this is insufficient to guarantee workload balance under the skewed key distributions, e.g., heavy tail distributions.

*c) Global Grouping:* data tuples are sent to a dedicated instance of a downstream operator, typically for results aggregation in the final stage of the topology. As an example, $TopK_1$ instance gathers tag counts from all instances of operator *Count*.

*d) All Grouping:* data tuples are duplicated and sent to all instances of the downstream operator, equivalent to data broadcasting.

*4) Instance Placement:* Given that user has configured parallelism information of the application operators in terms of instances, however, to realize this into practice, a framework scheduler will then accomplish this throughout a mapping between operator instances and physical computing machines which will host them. We refer to such mapping as an instance placement, which is usually carried out based on specific strategies including those simple ones such as random/round-robin assignment, or more sophisticated ones such as traffic-aware assignment [11], [18]. Once the instances are mapped and scheduled to the hosting machines, the data communication patterns between each pair of instances are fixed. For example, applying round-robin placement strategy to parallel version of our example application depicted in Figure 1b over M1, M2, M3, and M4 compute machines in Figure 2, yields to: M1$\leftarrow$ $Split_1, Job\ Extractor_1, Count_1$; M2$\leftarrow Split_2, Job\ Extractor_2, Count_2$; M3$\leftarrow Skill\ Extractor_1, Merge_1, TopK_1$; M4$\leftarrow Skill\ Extractor_2, Merge_2$.

### B. Network Model

Today's datacenters are built with the aid of rich interconnectivity such as Fat-Tree [19] and Leaf-Spine multi-rooted Clos [20], [21] modern network topologies (e.g., Figure 2), to support full-bisection bandwidth and to potentially make capacity of the fabric's internal links bottleneck-free. Also, the capacity of internal switch fabric (i.e., switching capacity) can support more than the capacity of all ports concurrently and the use of TCAM can even enable a high performance matching of high volumes of data packets. However, the main issue that remains affecting service delay and network utilization is the bottleneck due to the volume of traffic that could possibly overwhelm the capacity of outgoing ports.

In our analysis and implementation, we consider any datacenter fabric in which the bottleneck might occur at any link in the network. This is important because some flows might still happen to saturate some internal links and thus overwhelming switches buffers causing performance degradation of the application. Example case that might cause this issue is due to shortcomings of flow scheduling. Some flow scheduling algorithms to the network paths, such as ECMP algorithm, neither account for current link utilization nor flow volume. Subsequently, this makes capacity limit of the fabric the source of performance bottleneck.

Furthermore, the bottleneck more probably happens as a result of placement of application instances into the machines, which incurs some flows to mandatory traverse specific link to the destination instance. Therefore, the machines' directly connected links (i.e., uplinks and downlinks) become the largely potential places of bottleneck, and hence in our previous study, we primarily focused only on machines' directly connected links assuming the internal links of the fabric are always bottleneck-free and thus modeled the datacenter fabric as a resemblance to a big switch.

In this paper, we therefore consider a more generic network model taking into account that contention on the bandwidth might happen at any link in the network. With this model as depicted in Figure 2, we define the *uplink* (res. *downlink*) of each machine as its communication channels to (res. from) the rack switch. Also, we define the *rack-to-core* (res. *core-to-rack*) link of each rack (res. core) switch as its communication channels to the core (res. rack) switch. We use the term *internal link* for any of them.

Next, by optimally allocating the bandwidth of the internal links of the fabric in addition to the machines' uplinks and downlinks, we step forward general settings for better performance of application layer run on top of a multi-hop network.

### C. Communication Flows

We refer to a uni-directional data transfer between any given pair of instances as a flow, denoted by $f$. A flow is called an *internal* flow if its two communication instances are placed in the same machine, otherwise it is called an *external* flow. Recall in the the application we explained above about instance placement, the data transmission from $Split_1$ in M1 to $Job\ Extractor_1$ in M1 is an example of internal flow, while the flow from $Split_1$ in M1 to $Skill\ Extractor_1$ in M3 is an external flow.

As we adopt multi-hop network model, this leads to the fact that all the external flows traverse at least two uni-directional links. More importantly, when multiple flows are traversing the same bottleneck link, the data packets of the flows are queued in limited-size buffers connecting each consecutive nodes and hence packet dropping is most likely to happen due to the congestion. Thus, given limited-size buffers, the natural question to ask is how much bandwidth shall be allocated to each flow for its packet transfer on the uplink, downlink, and internal links over the multi-hop network. The key contribution of this paper is to share the bandwidth among these competing flows so as to maximize application-layer welfare. The welfare of stream application is realized by attaining low latency and high throughput.

### D. Motivation Example

In our preliminary study [22], we have measured and analyzed the impact and importance of bandwidth allocation on streaming applications. We have conducted a measurement study of different settings on a bandwidth-limited network carries streaming application traffic and have evaluated an application of 4 operators with parallelism set to 1 for each operator (Figure 3a). Specifically, we measured application-level throughput, as expressed in total number of completely processed tuples per second (averaged over 300-second experiments).

Figure 3b compares the overall throughput of the application under multiple placements, as denoted by TP1, TP2 and TP3. Under each placement strategy, we compared two different bandwidth allocation mechanisms, the default TCP congestion control (i.e., With TCP) and the best allocation (i.e., With BA) we obtained throughout brute-force search. We observed that a proper bandwidth allocation rather than vanilla congestion control can make further improvement on the overall application throughput. In particular, our observation is that instead of sharing the bandwidth based on max-min fair *rate* allocation as approximated by TCP, the proper sharing is to look for max-min fair *utility* allocation. Based on our analysis on the causes of performance degradation, we envision that utility function to account for application welfare, it must lead to that concurrent flows are processed altogether in a reasonable time window. Hence, utility functions should infer and do a favor for important and critical flows that otherwise application logic stalls data processing.

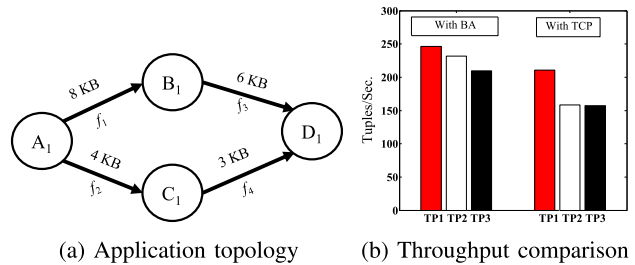

(a) Application topology   (b) Throughput comparison

Fig. 3. A motivation example: under three options for instance placements, we show the performance of bandwidth allocation (with TCP) versus the optimal (with BA).

Overall, in our example study in Figure 3, we found that the best bandwidth allocation has achieved 17%, 47%, and 33% improvement in placement TP1, TP2 and TP3, respectively. Nonetheless, brute-force search for the best allocation, e.g. as in [22], is too costly to be affordable in practice.

Therefore, in this paper, we rigorously develop max-min fair utility functions and use novel metrics of flow urgency.[1] for bandwidth, such that to entitle bandwidth allocation to the optimality as driven by application layer. Recall that such metrics are periodically retrieved such that to allow our optimization to work online over the lifetime of long-running stream analytics.

## III. BANDWIDTH ALLOCATION IN STREAMING APPLICATIONS

In this section, we first formulate a bandwidth allocation problem of streaming application, followed by a discussion of main challenges to resolving this problem.

### A. Problem Formulation

Consider a datacenter network that consists of a set $\mathcal{L} = \{1, ..., L\}$ of links of capacity $C_l$, $l \in \mathcal{L}$. The network is shared by a set $\mathcal{F} = \{1, ..., F\}$ of flows. We denote the rate of any network flow $f \in \mathcal{F}$ by $x_f$. Our goal is to find a vector $\mathbf{x} = (x_1, x_2, ..., x_F)$ of flow rates that maximizes the overall

---

[1]Flow urgency is not exactly its sender demand, but rather its relative demand to other flows according to the importance for the application.

application welfare $U(\cdot)$, s.t. constraints of link capacities:

$$\max_{\mathbf{x}} \quad U(\mathbf{x}) = U(x_1, x_2, \ldots, x_F) \tag{1}$$
$$\text{s.t.} \quad \mathbf{Rx} \leq \mathbf{C} \tag{1a}$$

where, $\mathbf{R}$ is a binary valued routing matrix, of which $\mathbf{R}(f,l) = 1$ if and only if flow $f$ traverses link $l$; and $\mathbf{C}$ is the vector of link capacities. The constraint (1a) means that the aggregate flow rate at any link $l$ cannot exceed its capacity $C_l$.

In our context, the overall application welfare $U$ can be quantified by the aggregate processing rate of the streaming application and is associated with managing bandwidth allocation among application's *network* flows deployed over a multi-hop network interconnecting a set $\mathcal{M} = \{1, ..., M\}$ of machines with a set $\mathcal{I} = \{c_1, c_2, \ldots, c_K\}$ of internal links. We denote the uplink starting with machine $i$ by $u_i$, the downlink ending with machine $j$ by $d_j$, and any of the internal links the flow traverses denoted by $c_k \in \mathcal{I}$. Furthermore, as the application involves active flows for unbounded time, hence the optimal control of flow rates might change over time. We therefore denote the vector $\mathbf{x}$ of flow rates at time $t$ by $\mathbf{x}(t)$ and re-describe the problem in (1) as to find the optimal $\mathbf{x}^*(t)$ at time $t$ that maximizes $U$, s.t. capacity constraints of machine uplinks and downlinks are satisfied.

$$\max_{\mathbf{x(t)}} \quad U(\mathbf{x}(t)) = U(x_1(t), x_2(t), \ldots, x_F(t)) \tag{2}$$

$$\text{s.t.} \quad \sum_{s(f)=i, d(f)\neq i} x_f(t) < C_{u_i}, \quad \forall i \in \mathcal{M}, \tag{2a}$$

$$\sum_{d(f)=j, s(f)\neq j} x_f(t) < C_{d_j}, \quad \forall j \in \mathcal{M}, \tag{2b}$$

$$\sum_{f \in \mathcal{F}_{c_k}} x_f(t) < C_{c_k}, \quad \forall c_k \in \mathcal{I} \tag{2c}$$

where, $s(f)$ and $d(f)$ denote the source and destination of flow $f$, and $\mathcal{F}_{c_k}$ denotes the set of flows share the internal link $c_k$. The constraints (2a), (2b), and (2c) ensure that the aggregate flow rates at any machine uplink $u_i$, downlink $d_j$, and internal link $c_k$ do not exceed the uplink capacity $C_{u_i}$, the downlink capacity $C_{d_j}$, and internal link capacity $C_{c_k}$ respectively. However, the key challenge is in defining non-clairvoyant flow utilities $U(\mathbf{x}(t))$ that through them network flows are allocated proportional rates to their volumes [22], without prior knowledge of flow volume; meanwhile, the flow volume should be estimated in the presence of continuously- and timely-varying load on processing pipeline. Before we show how to derive flow utility functions and how to apply bandwidth allocation based on them, we present following specific challenges to network flows in stream analytics.

### B. Challenges

- **Unbounded flows transfer**: A data stream is an unbounded sequence of events over time. This unbounded nature of data streams makes their corresponding network flows unbounded as well. Most existing state-of-the-art approaches are focused primarily on management of bounded network flows such as network flows of MapReduce jobs or search queries. For example, scheduling flows based on flow's remaining size [23], or bytes sent by each flow [24] or coflow [25] have been introduced to minimize completion time by mimicking the *Shortest Job First* (SJF) approach. However, these techniques remain hard to adapt to unbounded network flows of stream application.
- **Undetermined flows volume**: Also, the unpredictability and variability are common attributes of data stream sources. The variability in data streams makes it hard to obtain accurate flow information because flow volume in terms of tuple unit size and sending rate usually change with time. Consequently, bandwidth allocation policies that rely on prior knowledge of flows [23] or coflows [26] remain also inapplicable for stream applications. In result, bandwidth allocation policy has to capture flow updates and be adaptable to changes.

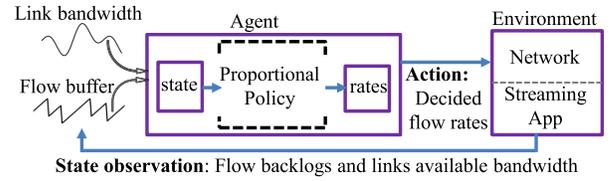

Fig. 4. The bandwidth allocation optimization framework for stream analytics.

Given these challenges, we introduce a cross-layer bandwidth control framework under which streaming application and network can be flexibly characterized to each other. We particularity show that bandwidth allocation policy we aspire to find, can be seamlessly transpired and materialized and in online manner.

## IV. BANDWIDTH ALLOCATION OPTIMIZATION FRAMEWORK

As we mentioned earlier, solving bandwidth allocation problem requires deriving subtle flow utilities that through them the uncertainty of application's traffic can be transpired and accordingly network flows are allocated proportionally. To address this problem, we propose an end-to-end *agent-environment* optimization framework, as shown in Figure 4.

In this framework, the agent observes a set of metrics, measured from the environment, including flow's state from the application layer and network links' available capacities from the network layer. We develop novel metrics of flow's state to determine flow utility in a non-clairvoyant manner that provides immediate insights of the urgency of the flow to the performance of the application layer. The agent then feeds these measured values to a bandwidth allocation algorithm, to compose flow utilities and to perform an optimization. Finally, the agent takes the result of the optimization as an action, by sending a rate vector to the network for regulating network flows in the next time interval. The entire process will be repeated alongside the lifetime of streaming application.

In the following, we describe the components of proposed bandwidth allocation optimization framework (Figure 4).

## A. Input Details

To control each flow rate, it is necessary to obtain at the first place the association relationships between data transmission flows and the underlying physical links. In streaming application, this can be achieved by sorting out the parallel version of application's logical topology, together with its placement map into the physical machines. Luckily, this information is easily to retrieve from master of the stream processing platform (e.g., Nimbus of Apache Storm) or from the cluster manager (e.g., Application Manager of YARN or Mesos in Apache Heron and Samza). Now that we know the the association between flows and physical links, the second step then is to record flows' state information and to determine links' available capacities. The state information are particularly intrinsic to characterize each flow volume in the presence of traffic variability in order to use them in estimating each flow utility. Meanwhile, available links capacity are required to optimize flow rates accordingly.

*1) Flow Buffer:* To discern what bandwidth is needed for each flow $f$ of unkown volume $V_f$ in streaming application, instead of using flow rate to express flow volume, we propose a *flow state* model to determine the state of the flow with changes through time. The reason of not using flow rate to express flow need is because rate observed presumably depends on cross-traffic in the network which does not reflect what bandwidth is really needed for this flow. Conversely, the state model can infer flow volume by profiling the actual data transfer of the flow and its status at endpoints (i.e., sender and receiver), as depicted in Figure 5.

To obtain actual data transfer of the flow, we keep recording amount of sent tuples at the application-level sender function, expressed in MB; meanwhile, for flow status, we keep track flow backlogs at flow endpoints via maintaining and measuring each endpoint with a dedicated queue. The sender endpoint backlog is computed as a queue length in MB of data tuples that are waiting in the queue for transfer service across network link. We use this metric to indicate how sending rate of the flow's sender endpoint is higher than available bandwidth for this flow. This case is used in each fork instance of the application. On contrary, the receiver endpoint backlog is computed as a queue length in MB of received data tuples that have been waiting in a queue for processing service. However, the latter case is used in each join instance of the application. For example, consider a receiving instance that performs join processing of tuples originating from flows of multiple sending instances. In such a case, the receiving instance might be *stalled* if some of the required flow's data tuples have not received yet due to network congestion. This leads to that flow of available tuples can not be further processed, and thereby not only this flow processing will be delayed, but also the entire pipeline processing. Due to also constant streaming of data tuples, we observed that the delay of flow processing overruns instance's memory capacity which might largely cause memory overflow (e.g., OOM). Therefore, we use queue length in this context to indicate the degree of instance stalling. Thus, to alleviate stalling, the unavailable flow's data tuples should be brought much faster by allocating it higher bandwidth than delayed flow.

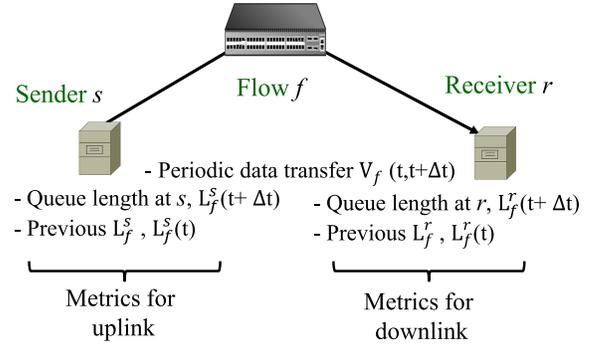

Fig. 5. A model of flow state.

The output of buffer profiling is a *5-metric* tuple characterizing state of each flow $f \in \mathcal{F}$ in a time interval $(t, t + \Delta_t)$, as depicted in Figure 5. The metrics include queue length of the flow $f$ at the sender and receiver at time $t$ denoted by $L_f^s(t)$ and $L_f^r(t)$; the queue length of the flow $f$ at the sender and receiver at time $t + \Delta_t$ denoted by $L_f^s(t + \Delta_t)$ and $L_f^r(t + \Delta_t)$, and the actual flow's data size transferred within $t$ and $t + \Delta_t$ denoted by $V_f(t, t + \Delta_t)$. A 3-metric tuple $< L_f^s(t), V_f(t, t + \Delta_t), L_f^s(t + \Delta_t) >$ is used for characterizing the flow state at the uplink and another 3-metric tuple $< L_f^r(t), V_f(t, t + \Delta_t), L_f^r(t + \Delta_t) >$ is used for characterizing the flow state at the downlink. For the flows contending for uplink, the higher values queue length of the flow, the higher demands of link bandwidth; while the flows contending for downlink the lower values queue length of the flow, the higher demands of link bandwidth. The profiling of flow states is thereby useful to compose flow utilities in order to serve each network flow need not only according to underlying network available capacity but rather based on flow importance to application performance. Shortly, we shall see how each flow state, expressed by the *5-metric* tuple, is used to compose flow utilities and to define optimization policy over these utilities that allocates bandwidth in proportional to each flow importance in the application performance.

*2) Link Bandwidth:* The bandwidth allocation algorithm has to know in addition to flow state, the network links allocatable capacities (i.e., available bandwidth). However, link allocatable capacity can be computed as a difference between its total capacity and the active flows rates over this link. Strikingly, in a cooperative SDN-based cluster in which a single administrative entity controls the network, all packets are transferred as instructed by forwarding rules stored in the network devices. This ensures that all network flows belong to multiple applications should be known over all machines links. Therefore, we leverage OpenFlow statistics features to collect flow statistics in order to estimate used bandwidth of each link. We use symbols $C_{u_i}$ and $C_{d_j}$ to correspond respectively to allocatable capacities of uplink $u_i$ and downlink $d_j$.

## B. Bandwidth Allocation Algorithm

In Algorithm 1, we summarize the end-to-end procedure of optimization of bandwidth allocation for improving the performance of stream processing application. The algorithm works in a time period ($\Delta_t$) basis to capture the change in the system. This allows the algorithm to work online in real-time to collect the measurement of flow state in the past immediate $\Delta_t$ and to respond by allocating flow rates for the next immediate $\Delta_t$. Given the network model in Section II-B in which machine' uplinks, downlinks, and the internal links (line 2) subject to their available bandwidth (line 7) are the resources of target for optimal use. Also, given that flow assignment over these links are determined (lines 4, 5, and 6 respectively) and each flow state (line 16) is also obtained based on the model in Figure 5, thus the mechanism of bandwidth allocation is as follows.

---

**Algorithm 1** Online Bandwidth Allocation

1: **Input**:
2: Uplinks $u = \{u_i : i \in \mathcal{M}\}$, downlinks $d = \{d_j : j \in \mathcal{M}\}$, internal links $\mathcal{I} = \{c_1, c_2, \ldots, c_K\}$
3: Network flows $\mathcal{F} = \{1, \ldots, F\}$
4: Machine's uplinks flow set $\{\mathcal{F}_{u_i} : i \in \mathcal{M}\}$
5: Machine's downlinks flow set $\{\mathcal{F}_{d_j} : j \in \mathcal{M}\}$
6: Internal links flow set $\{\mathcal{F}_{c_k} : c_k \in \mathcal{I}\}$
7: Allocatable capacities of links $\{C_l : l \in \mathcal{L}\}$
8: **Output**:
9: Proportional bandwidth fair share $x_f$ during next time interval $\Delta_t$ for each bottlenecked flow $f$
10: **Initialization**:
11: Bottlenecked uplinks, downlinks, internal links and flows are $\omega^b \subset u$, $\phi^b \subset d$, $\varphi^b \subset \mathcal{I}$ and $\mathcal{F}^b \subset \mathcal{F}$ respectively;
12: $t \leftarrow 0$;
13: $L_f^s(t) \leftarrow 0$, $L_f^r(t) \leftarrow 0$, $\forall f \in \mathcal{F}^b$
14: **do**
15: run the streaming system for time $\Delta_t$;
16: record $V_f(t, t+\Delta_t)$, $L_f^s(t+\Delta_t)$ and $L_f^r(t+\Delta_t)$, $\forall f \in \mathcal{F}^b$
17: **for** each uplink $u_i \in \omega^b$ **do**
18:    Solve the optimization problem (3)
19: **end for**
20: **for** each downlink $d_j \in \phi^b$ **do**
21:    Solve the optimization problem (4)
22:    $x_f(t+\Delta_t) = \min\{(x_f^u(t+\Delta_t), (x_f^d(t+\Delta_t)\}$.
23: **end for**
24: **for** each internal link $c_k \in \varphi^b$ **do**
25:    $D(c_k) = \sum_{f \in \mathcal{F}_{c_k}} x_f(t+\Delta_t)$
26:    **if** $D(c_k) > C_{c_k}$ **then**
27:       $x_f(t+\Delta_t) = \frac{x_f(t+\Delta_t) C_{c_k}}{D(c_k)}$, $\forall f \in \mathcal{F}_{c_k}$
28: **end for**
29: $x_f(t+\Delta_t) = \min\{x_f(t+\Delta_t)\}, \forall c_k \ni f$
30: $t \leftarrow t + \Delta_t$
31: **While** ( $\exists f \mid f \in \mathcal{F}^b \wedge (L_f^s(t) \neq 0 \vee L_f^r(t) \neq 0)$ )

---

As we mentioned earlier, from our analysis of mapping of application's flows into machines uplinks and downlinks, we observed that stream applications follow a *Fork-Join* like communication pattern in common. In the *Fork* stage, network flows of instance(s) co-located at the same machine compete(s) for machine's uplink, while in the *Join* stage, the instance(s) at certain machine, receive(s) multiple flows from some other instances that compete for machine's downlink. One more important observation is that majority of streaming applications flows sharing network bandwidth have to be concurrent within a reasonable time window in order to be processed together. We use these two observations to derive bandwidth allocation algorithm to maximize the aggregate processing rate of streaming application, based on optimization problem (2). Therefore, to achieve this, the algorithm should make sure that if the input speed of data generated by the sender machines keeps unchanged during the next period of time, so as to assist the system processing all flow backlogs in all machines in the shortest time.

$$\min_{x_f^u(t+\Delta_t)} \max_{f \in \mathcal{F}_{u_i}} \frac{V_f(t, t+\Delta_t) + 2L_f^s(t+\Delta_t) - L_f^s(t)}{x_f^u(t+\Delta_t)} \quad (3)$$

$$\text{s.t.} \sum_{f \in \mathcal{F}_{u_i}} x_f^u(t+\Delta_t) = C_{u_i}, x_f^u(t+\Delta_t) \geq 0 \quad (3a)$$

$$\min_{x_f^d(t+\Delta_t)} \max_{f \in \mathcal{F}_{d_j}} \frac{L_f^r(t+\Delta_t) + x_f^d(t+\Delta_t)\Delta_t}{[V_f(t, t+\Delta_t) - L_f^r(t+\Delta_t) + L_f^r(t)]/\Delta_t} \quad (4)$$

$$\text{s.t.} \sum_{f \in \mathcal{F}_{d_j}} x_f^d(t+\Delta_t) = C_{d_j}, x_f^d(t+\Delta_t) \geq 0 \quad (4a)$$

Next, we use this idea to derive time-utility functions that express the time urgency of each flow. Based on them, we explain the derivation of the optimization problems (3) and (4) of our algorithm. If an uplink $u_i$ is shared by multiple flows whose set is denoted by $F_{u_i}$, then it is under the *Fork* pattern. Thus to define flow utility for any flows $f \in F_{u_i}$, the data amount of the flow $f$ generated by the sender machine $i$ during the time interval $(t, t+\Delta_t)$ is $V_f(t, t+\Delta_t) + L_f^s(t+\Delta_t) - L_f^s(t)$, i.e., the data size $V_f(t, t+\Delta_t)$ of the flow $f$ transferred plus the variation $L_f^s(t+\Delta_t) - L_f^s(t)$ of the queue length of the flow $f$ in the sender machine $i$ during the time interval $(t, t+\Delta_t)$. If the generating speed of the data of the flow $f$ keeps unchanged during the next time interval $(t+\Delta_t, t+2\Delta_t)$, then there will exist the data of the size $V_f(t, t+\Delta_t) + 2L_f^s(t+\Delta_t) - L_f^s(t)$ needed to be transferred during the time interval $(t+\Delta_t, t+2\Delta_t)$. Under the bandwidth $x_f^u(t+\Delta_t)$ allocated to the flow $f$ during the time interval $(t+\Delta_t, t+2\Delta_t)$, the total time to finish transferring the data is at least $\left[V_f(t, t+\Delta_t) + 2L_f^s(t+\Delta_t) - L_f^s(t)\right]/x_f^u(t+\Delta_t)$. Because our goal is to make the system finish processing all backlogs of the flows sharing the link $i$ in the shortest time, which usually occurs when the maximum of the transferring times $\left[V_f(t, t+\Delta_t) + 2L_f^s(t+\Delta_t) - L_f^s(t)\right]/x_f^u(t+\Delta_t)$ among the flows of sharing the uplink $i$ is minimized, we derive the optimization problem (3). Based on a similar idea, we can also derive the optimization problem (4).

We empirically measured how different values of $\Delta_t$ in optimization problem (4) affect rate allocation and observed that $\Delta_t$ slightly affecting the rate allocation decisions and accordingly the transferring times. The mean accuracy is more than 99% and with standard deviation of 0.02. Thus, $\Delta_t$ remains a constant factor among all flows. Conversely, flows' state metrics keep accumulating with respect to the time during $\Delta_t$, hence it is conceivable that these metrics differentiate the flow's necessity for the bandwidth. Furthermore, it is worthy to note that $\Delta_t$ as event (not as parameter in Equation 4) to inquiry and update the system depends on the the magnitude of the time variation of the load on the analytics pipeline. If the flow's volume changes rapidly, the $\Delta_t$ is better to be relatively small such that to capture the changes on the flow state quickly and to update the system with the new optimal rate allocation.

Having solved optimization problems (3) and (4), we succinctly reveal problem's abstract formulation (2) and fulfill its main goal. The constraints, (3a) ensures that aggregate of allocated flow rates at any machine uplink $u_i$ does not exceed uplink capacity $C_{u_i}$, and (4a) ensures that aggregate of allocated flow rates at any machine downlink $d_j$ does not exceed downlink capacity $C_{d_i}$. The symbols $x_u$ and $x_d$ are per-flow $f$ allocated rate at flow's uplink and downlink, respectively, and the flow will be given the minimum allocated rate $x_f$ of either (line 22). Besides, if the data flow traverses any of the congested internal links, then the flow will be allocated proportional rate (line 27) to its informed rate (line 22) in the uplink and downlink. In case the flow shares multiple bottleneck internal links, then it will be allocated the minimum rate among them (line 29).

Overall, when network flows share bottleneck uplink, downlink, or internal links are unequal in their volumes (i.e., estimated by flow state) and are required to be processed concurrently (i.e., time urgency), the default transport (e.g., TCP) in processing frameworks is obviously ill-suited because it is unaware of the application atop how such variabilities impact its performance. On contrary, our algorithm checks state of each flow at each congested uplink, downlink, and internal links and rigorously solves optimization problems to distill the optimal rate for each flow over respective links (lines 10-29). This continues alongside any of the flow buffers are not empty (line 31) and in each time interval outputs per-flow optimal allocation.

## V. IMPLEMENTATION

In this section, we present the details of integrating SDN into analytics platform and how cross-layer information can be exchanged in an automated and flexible manner so as to provide a high-performance analytics pipeline. To accomplish this, we translate our optimization framework described in Section IV into a multi-tier architecture (Figure 6): Streaming application management plane (tier 1), SDN-based control plane (tier 2), and SDN-based forwarding devices plane (tier 3). We materialize streaming application manager in Storm framework [2], the popular open-source stream processing platform. Meanwhile, we implement SDN-based control in OpenDaylight SDN controller [16], the largest community-led and industry-supported open source SDN framework. For the SDN-based forwarding plane, this is the tier wherein network bandwidth resources reside and in need of effective management to maximize streaming application performance. In tier 3, to interoperate with above tiers, we adopt SDN-based devices which dispense with arriving data packets of streaming application according to the desired performance goals elaborated by the management plane. Taken together, the streaming application manager in tier 1 will then quickly and dynamically program the network devices in tier 3 via some of the abstract interfaces supported by network controller in tier 2, as follows.

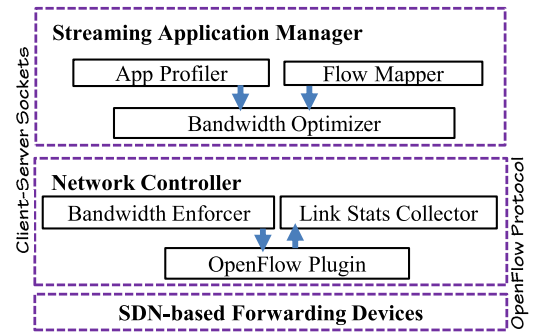

Fig. 6. SDN-based implementation of proposed bandwidth allocation algorithm.

### A. Tier 1: Streaming Application Manager

This tier is the core of our solution, it implements the agent decision-maker in above agent-environment framework. By observing the state of all network flows encompass streaming application and the available links bandwidth, streaming application manager then solves optimization problems based on proposed bandwidth allocation algorithm and provides the optimal rates allocated to network flows for next time period. The main components of this manager are summarized as follows.

*1) Flow Mapper:* To determine the optimal bandwidth for each flow, bandwidth allocation algorithm should know as a first step application flows to network links mapping over all machine uplinks, downlinks, and internal links. We built a customized scheduler instead of Storm default scheduler in order to maintain a deterministic map of streaming application's instances over the compute machines.

*2) Application Profiler:* This entity is also known as flow buffer, it aims to monitoring and collecting the state of all network flows to feed them as input to the bandwidth optimizer component. Driven by our model of flow state, application profiler keep tracks each flow in the application and records its amount of transferred data (MB) and backlog (MB) at both flow's receive queue of received messages and flow's send queue of sent messages, is depicted in Figure 5.

*3) Bandwidth optimizer:* This module is responsible for execution of bandwidth optimization algorithm. At each time interval, it pulls the application and the network to get the inputs and sends the outputs of the algorithm. Specifically, the optimizer collects the input from flow mapper, application profiler, and link stats collector to execute optimization algorithm and subsequently sends the output of the algorithm (i.e., optimal flow rates) to the network controller.

For this purpose, we develop a socket-based client-server interface to enable 2-way communication between bandwidth optimizer resides at tier 1 (i.e., at application layer) and bandwidth enforcer resides in SDN-based control at tier 2 (i.e., at network layer). Conversely, for the modules co-located in the same tier, they can access each others via local APIs.

## B. Tier 2: Network Controller

This tier is implemented in the world's largest open source SDN platform, an OpenDaylight (ODL) controller, which acts as middleware that orchestrates and facilitates exchange of control messages between application manager (tier 1) and Openflow-based physical network (tier 3). For our purpose, in addition to the core components of ODL controller, we develop a native MD-SAL compliant network application within the controller to primarily implement bandwidth allocation and to collect link measurements and statistics information. In particular, this application consists of bandwidth enforcer and link statistics collector modules. Technically speaking, our development of such a native application in ODL encompass different technologies such as OSGI, Karaf, YANG modeling, blueprint container, and a set of distinctive messaging patterns including RPC, publish-subscribe, and datastores accesses [16]. We skip the modular design and implementation details for the sake of brevity.

*1) Bandwidth Enforcer:* This module is built to dynamically update the rates of network flows according to the output of optimization algorithm. Instead of directly modifying the underlying transport mechanisms (such as TCP), we leverage the metering API feature supported in OpenFlow-based networks to enforce the output rates received by bandwidth optimizer into the datapath of the underlying network devices. Specifically, we associate each network flow with a meter under which the packets belonging to this flow are allowed to pass through the egress ports of the switch for up to specific upper-bound rate (e.g., the rate decided by bandwidth optimizer). Subsequently, the bandwidth enforcer instructs OpenFlow plugin to translate allocated rates into OpenFlow messages and to install them into the networking devices.

*2) Link Statistics Collector:* We implement this module to estimate the available bandwidth of the link particularly when the network is injected with cross traffic from multiple applications. In OpenFlow network, all packets belong to network flows are transferred based on flow forwarding rules stored in the network devices. Meanwhile, we register all these rules with statistics service module in ODL. The module then uses the service APIs implemented by OpenFlow plugin to send statistics requests to network devices to report flow statistics including packets count, bytes count, and duration.

*3) OpenFlow Plugin:* This component is one of the essential components in ODL which implements OpenFlow protocol to mediate the communication between underlying network devices and network control applications. It is used in tandem with functions of network application developed in tier 2 to interact with the underlying network devices supporting OpenFlow protocol.

## C. Tier 3: SDN-Based Forwarding Devices

This tier is primarily the data plane of the network interconnecting machines using SDN-enabled devices that can be programmed directly by the centralized network controller resides in tier 2, using OpenFlow protocol. Each device contains a pipeline of flow tables used to store flow rules installed by the controller under which the main set of network functions such as forwarding, rate control, and routing are supported. More recently, some SDN-enabled devices also support a meter table to store meter entries to implement bandwidth-limiting in the ingress processing of received packets. The meter works like a token bucket policer which measures and controls the rate of packets belongs to each flow for up to specific pre-defined rate (e.g., the rate advised by bandwidth optimizer).

## VI. EXPERIMENTAL EVALUATION

In this section, we evaluate the performance of our proposed bandwidth allocation model with a real hardware testbed. Our testbed experiments illustrate proposed model's good performance for realistic and synthetic workloads of real streaming applications. Our detailed experiments confirm that our model also achieves close-to-TCP in terms of network utilization.

### A. Setup

*1) Testbed Experiments:* We use brocade ICX-6610 24-port Gigabit SDN-enabled hardware switch with 1Gbps uplinks and downlinks to test proposed optimizing algorithm in a non-blocking setting where the congestion is only restricted to machine' directly connected links. Furthermore, to evaluate the proposed optimization particulars in a more general setting with multi-hop network, we built a fat-tree like testbed of 7 switches, as shown in Fig. 2(c). In this setting, we throttle the internal links capacities such that to shift the bottleneck from the machines' uplinks and downlinks to the internal links of the testbed.

In both aforementioned settings, we set up 10-machine Storm cluster, each of 4-core Intel Xeon E5-1620 3.5GHz CPUs, 16GB of RAM, and 1TB HDDs. One machine is configured to run Storm master (known as *nimbus*), Zookeeper, and bandwidth optimizer. Eight machines are designated for Storm worker nodes (known as *supervisors*) to run the instances of experimental streaming applications. During experiments, the worker nodes were kept in sync by using the standard network time protocol (NTP) on the Ubuntu Linux. We also use one machine to host OpenDaylight SDN controller including our proposed bandwidth enforcer and statistics control plane applications.

*2) Test Applications and Benchmarks:* We have implemented two real-time stream analytics applications on top of Apache Storm: Trending Topics (TT) and Trucking IoT (TI). Figure 7 shows the topologies of the TT and TI applications. TT application considers a topic as trending when it has been among the top $K$ topics in a given window of time. We implement a topology of this application which consists of chain of five operators, in which each operator has one or more instances. The first operator is source of the topology that emits unbounded sequence of data streams to next operator to split them into words and emits them to the next operator, word count operator (WCT), to perform and maintain word counting. Then, WCT waits a time equals to $K$ arrivals and performs emitting to an aggregator operator. The latter will jointly process the receiving tuples from multiple WCT instances, extracts the top $K$ trending topics from all of them, and emits the results to report operator. For TT, we use a real dataset from twitter which contains millions of tweets for about 3 years. We set the sliding window to 30 seconds and emulate average arrival rate of 1000 tweets per second.

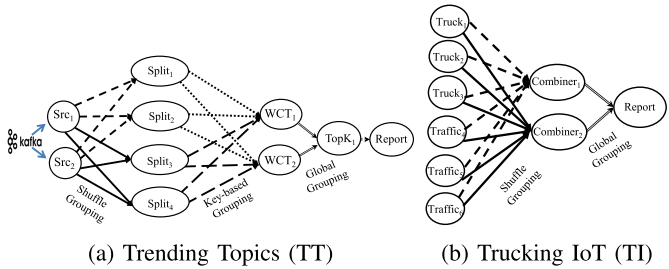

Fig. 7. Topologies of two test streaming applications.

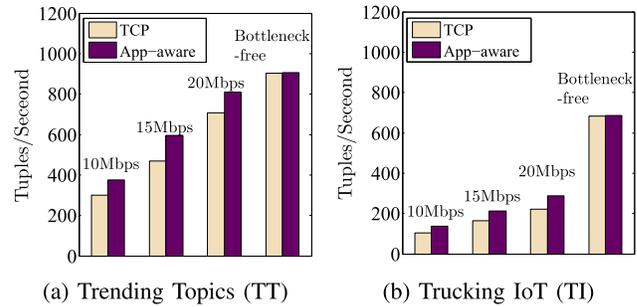

Fig. 8. Application Throughput (in uplink/downlink bottleneck).

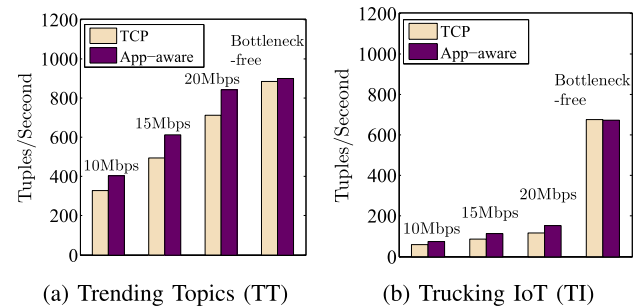

Fig. 9. Application Throughput (in a multi-hop bottleneck).

The TI application, is the widely used one in its topology design by many of streaming applications such as frequent pattern detection [13] and distributed computer-vision pipelines [27]. Trucking IoT service performs real-time analysis on IoT data streams coming from multiple sources. This topology receives two different streams, one stream reports data about truck status while the other one about traffic congestion status. The application processes both streams concurrently in a way that data from the truck is combined with the most up-to-date congestion data and reports a timely action that should happen accordingly. For TI, we use two synthetic datasets of different data tuple sizes which emulate the sizes of real tuple sizes reported by each of the truck sensor and traffic congestion online source service. We also emulate the arrival rate of 250 tuples per second of each stream.

With the TT and TI applications, we evaluate the performance of streaming application when the network bandwidth is the bottleneck possibly, i.e., the derived tuples rate from the application or data stream ingestion rate is higher than available network bandwidth, respectively.

*3) Baseline:* We use the standard TCP bandwidth allocation model as our baseline. It is the default model used by streaming frameworks like Storm [2], Heron [3], and Flink [5]. We compare TCP against our proposed bandwidth allocation for streaming applications, which we call it, *App-ware*, for purpose of identification.

We ran a series of experiments each with 600 seconds. In all experiments, we set a sampling rate for the per flow's 5-metric tuple to 1 per second in order to report flows backlog to application profiler. Also, we set the timer interval $\Delta_t$ to 5 seconds to periodically perform new bandwidth allocation by bandwidth optimizer. We repeat each experiment 4 times over the cluster of different available bandwidth. In particular, we set link bandwidth to 10Mbps, 15Mbps, and 20Mbps to evaluate impact of different bandwidth bottlenecks for running application. We also run the applications on the cluster with bottleneck-free setting (i.e., sufficient available capacity).

### B. Performance Improvement

For evaluating the performance of streaming application, we use the widely used metrics of interest by streaming frameworks, *application throughput* and *average end-to-end latency*. Application throughput is the average number of successfully processed tuples per unit time by the sink operator of streaming application, while average end-to-end latency is the average time taken over all tuples from the point each tuple leaves the source until it gets completely processed by sink operator.

*a) Application Throughput:* Figure 8 contrasts throughput of TT and TI based on App-aware versus TCP over each of 10Mbps, 15Mbps and 20Mbps settings, in which the link bottleneck is restricted only to the machine' directly connected links (i.e., uplinks and downlinks). The experimental results confirm that App-aware outperforms TCP by 25%, 27%, and 15% in TT, and by 30.93%, 30.27%, and 30.80% in TI. Also, Figure 9 illustrates similar results but in the multi-hop network bottleneck. In particular, the experimental results confirm that App-aware outperforms TCP by 19.21%, 19.52%, and 15.69% in TT, and by 22.09%, 23.09%, and 23.98% in TI.

*b) End-to-End Latency:* Figure 10 contrasts average end-to-end latency of TT and TI based on App-aware versus TCP over each of 10Mbps, 15Mbps and 20Mbps settings in the restricted bottlenck setting. The experimental results confirm that App-aware outperforms TCP by 14.27%, 26.24%, and 50.17% in TT and by 5.72%, 10.10%, and 17.28% in TI. Furthermore, Figure 11 shows partially similar results but in the multi-hop network bottleneck setting. In particular, the experimental results confirm that App-aware outperforms TCP by 11.20%, 15.90%, and 60.89% in TT, but the latency results in TI almost look similar in both. The reason is actually because the internal links are heavily congested, and hence the data packets experiencing the worst delays which might be preferentially dropped or last long time in the switches buffer.

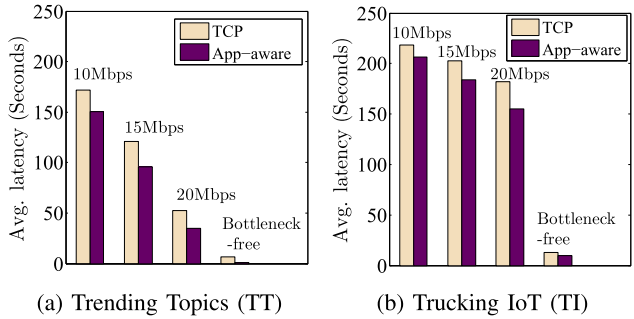

Fig. 10. End-to-end Latency (in uplink/downlink bottleneck).

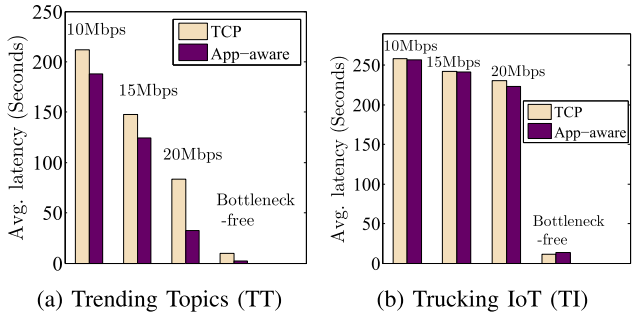

Fig. 11. End-to-end Latency (in a multi-hop bottleneck).

In link bottleneck setting, the improvement in application throughput and end-to-end latency can be interpreted by the fact that App-aware strives to allocate bandwidth to the flows proportionally to their importance (e.g., time urgency for window-based join processing) in application performance rather than based on bandwidth fairness as what TCP does. In TI, each combiner instance requires existence of data tuples from both source instances. Therefore, in TCP large data tuples flows often get throttled by some other very frequent small data tuples flows which leads to processing stall of the combiner. In contrary, App-aware relies on flow metrics and then smartly allocates the flows proportional bandwidth, which in turns alleviates instance's stalled time and speeds up needed data tuples arrival. For TT, key-based grouping along with accumulating top $K$ word at the WCT instances create flows with unbalanced sizes. Those flows are required by TopK aggregator to decide the final Top $K$. Hence, TCP falls short to express such imbalance, while App-aware captures such imbalance and allocates the proper bandwidth which greatly helps application achieves better performance. In bottlenck-free setting, for both applications, the performance of App-aware is much similar to TCP and sometimes it performs better.

### C. Link Utilization

Besides, link utilization is an important property for any bandwidth allocation algorithm to use the entire available bandwidth. To evaluate network utilization due to dynamic bandwidth allocation by our model, we use *average link throughput*, the average of aggregate throughput over all bottlenecked links in the cluster.

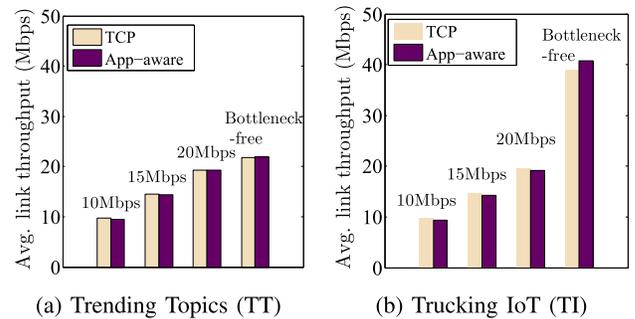

Fig. 12. Link utilization.

Though TCP congestion mechanism is application-agnostic, it utilizes bottleneck links very well. Figure 12 contrasts average link throughput of TT and TI as results of bandwidth allocation based on App-aware and TCP model. App-aware provides an average of 99% and 97% link utilization much the same as TCP link utilization for TT and TI, respectively. The model utilizes all available bandwidth and in case some bandwidth remains, the model performs backfilling pass to allocate the remaining bandwidth among the flows according to their proportional share in the previous pass.

### D. Overhead of Bandwidth Optimizer and Enforcer

In our solution framework, the switches periodically send/receive updates to/from the ODL network controller. Although the timing interaction between the controller and the switches is important. However, in our solution of stream application running for unbounded time, the interaction requirement is not in milliseconds or microseconds scale, but rather in seconds scale. Thus, the majority of the packets are parsed in the data-plane without going through ODL and only a periodic update of the available bandwidth and decisions of flow rate are exchanged. It is thus why the processing load for both switches and controller due to our design is not much heavy.

Specifically, the most important components that influence solution decision of the whole framework are computation and communication overhead of bandwidth optimizer and enforcer, respectively. To evaluate the computation overhead of bandwidth optimizer, we report CPU time at time step of new allocation in Trucking IoT application running for 600 seconds. We observed that optimizer took 6 milliseconds on average to extract flow statistics and to calculate the optimal bandwidth allocation. Further, to evaluate the communication overhead of bandwidth enforcer, we log the completion time of each flow rate update at one time step of the new allocation. We observed that time for the controller to completely update the switch with new flow's rate (i.e., meter table update) ranges from 100s of microseconds to 10s of milliseconds. As a result, such timing's overhead is negligible based on the interaction requirement of our model and therefore is able to cope with dynamic changes during application optimization.

## VII. FURTHER EXTENSION FOR MULTIPLE APPLICATIONS

App-aware mechanism that we have introduced ensures a utility fairness according to urgency of the flow to the performance of the application layer. This mechanism is mainly focused on the performance improvement of an individual real-time streaming application, however, it is likely that multiple streaming applications and cross-traffic that might also belong to different users can coexist in the same cluster. Such heterogeneity leads to fundamental tradeoffs in network resource allocation which might possibly lead to unpredictable performance for both users and cloud network providers [28]. Thus, it is still an open question as how to investigate performance particulars to best regulate sharing of network resources among multiple streaming applications with varied performance objectives and in the presence of cross-traffic.

As mentioned at the beginning of the paper, when the link becomes congested many of TCP variants approximate max-min bandwidth fairness on a flow-basis, hence it is common practice that application of many flows is likely to receive higher portion of bandwidth than others. In [28], theoretical allocation policies have been introduced as starting points in navigating the tradeoffs space of bandwidth allocation in cloud network, but remain of need to practical deployment. Here, we make a step forward and discuss how tier 2 of our solution framework can be leveraged to schedule multiple streaming applications that might have different degree of sensitivity to the bandwidth allocation. Such differences can be declared by defining different utility functions and are optimized by defining appropriate bandwidth allocation mechanism. In this space, we implement a point solution by assuming that optimal solution to the utility functions corresponds to the fairness of bandwidth allocation between the competing applications. That is, ensuring max-min fair allocation not only between the flows as what TCP can approximate but also between the applications as whole, regardless of number of flows belong to each application. An approach to implement this solution is by grouping applications into clusters of different priorities and by giving higher priority to applications of low achieved throughput. Similar applications in achieved average throughput $\mu^{(t)}$ up to time $t$ or in past immediate throughput $\mu^{(\Delta_t)}$ during period $\Delta_t$ could share a portion of link bandwidth collectively. Thus, we measure the throughput of an application $i$ at time $(t + \Delta_t)$ as given by:

$$\mu_i^{(t+\Delta_t)} \leftarrow \alpha \times \mu_i^{(t)} + (1-\alpha) \times \mu_i^{(\Delta_t)}, \quad (5)$$

where $\alpha$ is a constant weighting factor between 0 and 1.

*c) Implemenation of Group Scheduling:* We use *Link Statistics Collector* of our framework in Figure 6 to periodically record statistics of all flows belong to the active applications. Based on the idea of Equation 5, the groups are then identified with a simple clustering technique. Thereafter, the clusters are ready to be assigned priorities to allocate bandwidth that deserve. We assign group with lowest average throughput the highest priority and implement group priorities with a multi-level strict priority scheduler among a set $\mathcal{Q} = \{q_1, \ldots, q_m\}$ of queues, where $m = 8$ in our testbed. We associate the flows from applications belong to the same group into in same queue index among all egress ports. To control downlinks and uplinks, the queues for controlling downlink are readily available at egress port's in the hardware OpenFlow switches, while for uplinks the use of Open vSwitch (OVS) queues comes in handy to control the flows sharing the uplink. Both of them, namely the queues attached to downlinks and uplinks, can be easily programmed online via SDN controller so as to dynamically associate the flows to them based on scheduling algorithm.

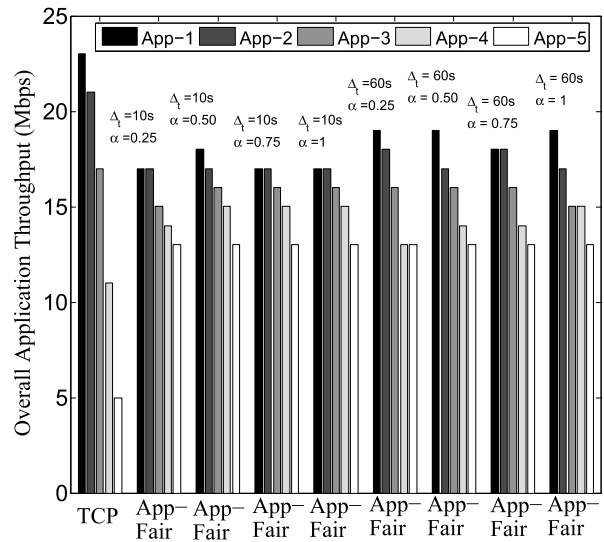

Fig. 13. Application-level Fairness.

To avoid starvation among the applications and to ensure that each application receives a non-zero bandwidth, the algorithm uses the periodic time-based measure of average throughput and signifies that to be fair the application should be displaced from one group to another in the next immediate schedule.

We evaluate the performance of the described mechanism, by experimentation of five competing applications for 10 minutes on the same testbed described in Section VI-A.1. To determine whether applications are receiving a fair share of bandwidth, we contrast TCP and our mechanism toward application-level fairness, named App-Fair in Figure 13, under differen values of $\Delta_t$ and $\alpha$. In TCP, the aggregate throughput of each application, App-$i$, where $i$ is number of flows belong to each application, is proportional to number of flows, while App-Fair attempts to fairly share the bandwidth among the applications regardless of their number of flows. Based on Jain index as a fairness measure [29], the preliminary results show that App-Fair is fairer than TCP. In particular, the fairness index of TCP based allocation is 84%, while App-Fair fairness index equlas 98%, 99%, 99%, and 98% at $\alpha$ equals 0.25, 0.50, 0.75, and 1 respectively and $\Delta_t = 10$ seconds.

Furthermore, it is important to note that grouping mechanism can be adapted to provide any desirable fairness via differential bandwidth allocation to meet some preference criteria.

## VIII. RELATED WORK

### A. Scheduling and Management in Datacenter

There has been a plethora of recent work on scheduling and management of tasks of datacenter applications with various performance objectives. Most of them aim to minimize completion time of flows belong to user-facing applications (e.g., web search) [23], [24], while others aim to minimize the completion time of data-intensive jobs (known as coflows, e.g., MapReduce jobs) as whole [25], [26], [30]. In principle, the basis behind introducing these solutions is driven based on the intuition of segregation (co)flows into short and long (co)flows, and then schedule each of them, broadly speaking, on a SJF basis. However, this kind of classification can not be directly applied for those applications do not have transfer boundary (e.g., streaming applications). On contrary, one aspect of our model design is addressed to overcome this limitation. It has the ability to inform the state of the flow based on its current and historic attributes regardless of its length. Moreover, recent research efforts focus more on sophisticated objectives such as scheduling mix-flows in commodity datacenters [31], or supporting bandwidth allocation of different service-level objectives ignoring application layer objectives [32].

### B. SDN-Based Traffic Management

Much like typical programming languages offer APIs to manage workload of computation resources, SDN interestingly offers APIs (e.g., OpenFlow) to manage workload of networking resources. Hence, SDN-based traffic management [33] has been adopted to enable efficient and dynamic management of network resources in datacenters during runtime of the applications. Hedera [34] and MicroTE [35] manage network flows using centralized network-wide scheduler in order to increase network throughput. Recently, Alkaff et al. [36] proposed to adopt SDN-based traffic management for optimizing cloud applications. They focused on the coordination between the application layer's task placement with network layer's route/path selection strategies. To our best knowledge, no recent work integrates SDN with real-time distributed streaming analytics to cooperatively optimize network bandwidth allocation for achieving the application-level performance requirements. Further, Wang et al. [37], investigate potential of integrating SDN controller with big-data application to facilitate more informed scheduling and placement decisions. Xiong et al [38] propose an SDN-based framework to improve the performance of queries over distributed relational databases.

### C. Resources Auto-Scaling in Stream Applications

DRS [13] has been introduced to schedule and provision computation resources to meet a real-time constraints.

SnailTrial [39] has been recently proposed to determine the importance of an execution activity in the transient critical paths of computation pipeline and to provide an immediate feedback for applications to perform automatic reconfiguration, dynamic scaling, or adaptive scheduling. DS2 [40] is a scaling controller for distributed streaming has been introduced to maximize system throughput via estimating the optimal level of parallelism for each operator. While these proposals are significant towards improving the performance of streaming pipeline, on contrary, our work address I/O-bound stream applications and introduce a bandwidth scaling model that is able to dynamically increase or decrease bandwidth allocation on a performance-centric basis of stream application. To the best of our knowledge, no recent work addresses network bandwidth allocation matter in streaming applications.

### D. Traffic-Ware Placement in Stream Applications

Several proposals [11], [12], [41], [42] have been proposed to avoid network transfers as much as possible. They aim primarily to collocate application instances in a few machines in order to minimize inter-machine communication. However, while these solutions to some extent improve the performance of the application, it is inevitable to distribute the applications into many more machines to not overload the CPUs of particular machines. Further, Typhoon [9] has recently presented an approach to optimize broadcast transfer pattern, but it is rather limited to specific pattern and further difficult to be applied because it requires modifying the routing table per instance involved in the broadcast transfer.

Nonetheless, optimization of bandwidth allocation is orthogonal to traffic-ware solutions. In our measurement-based study [22], we have shown that optimal placement is not sufficient either, but rather effective bandwidth allocation alongside optimal placement is required to ensure further optimal application-level performance.

### E. Network Utility Maximization (NUM)

A long line of optimization frameworks [43] began with Kelly's seminal paper [44] have been proposed for resource allocation based on NUM. While several of these frameworks have also been generalized to implement various flow utility functions such as rate, delay, jitter, and reliability [45], and recently NUMFabric [32] for different objectives such as weighted flow fairness, and minimizing flow completion times, we believe that none of them has capabilities to optimize application layer specific objectives to streaming analytics.

## IX. CONCLUSION

As we have observed that TCP-based bandwidth allocation of bottleneck links largely hurts application-level performance of streaming applications. As opposed to TCP and its variants, we have introduced a novel bandwidth allocation model that performs well with awareness of the application layer performance requirements. To make the proposed model practical, we have developed a cross-layer SDN-based framework which utilizes smartly the instantaneous information obtained from the application layer and provides on-the-fly and dynamic bandwidth allocation during the runtime of the streaming applications.

We have thoroughly investigated the performance of proposed solution through a series of testbed experiments with real-world stream analytics. The results reveal that application's performance resulted from our solution outperforms the standard TCP-based bandwidth allocation employed by most of streaming frameworks, and with a negligible overhead. It also performs comparable to TCP in network utilization. We believe that proposed model can be used not only by streaming analytics, but it can also be employed by several platforms involving parallel and pipelined network flows.

In addition, we have introduced an exemplary mechanism that leverages SDN framework to assist in sharing of bandwidth and reasoning of performance among multiple applications and we show a case for approximating application-level fairness. However, much more remains to be done in exploring the tradeoffs in bandwidth allocation for multiple streaming applications with varied performance objectives and in the presence of cross-traffic.

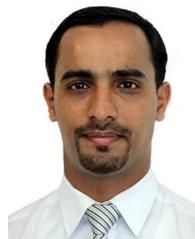

**Walid Aljoby** received the B.Sc. degree (Hons.) in computer information systems from Hashemite University in 2008 and the M.Sc. degree in computer engineering from the Jordan University of Science and Technology (JUST) in 2013. He is currently pursuing the Ph.D. degree with the Department of Computer Science, National University of Singapore (NUS). He is also a Research Scholar with the Advanced Digital Sciences Center (ADSC), the research center of the University of Illinois at Urbana–Champaign (UIUC), Singapore. His research interests include software-defined networking, cloud computing, real-time distributed stream analytics, congestion control, and security.

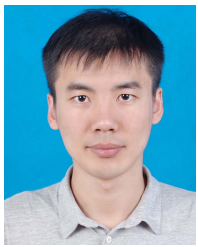

**Xin Wang** received the B.Eng. degree in computer science and technology and the Ph.D. degree in computer software and theory from the University of Science and Technology of China (USTC), in 2012 and 2017, respectively.
He is currently a Research Fellow with the Department of Computer Science, National University of Singapore. His current research interests include distributed systems and Internet economics.

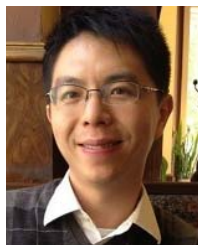

**Richard T. B. Ma** (SM'16) received the B.Sc. degree (Hons.) in computer science and the M.Phil. degree in computer science and engineering from The Chinese University of Hong Kong, in 2002 and 2004, respectively, and the Ph.D. degree in electrical engineering from Columbia University in 2010. During the Ph.D. degree, he was a Research Intern with the IBM Thomas J. Watson Research Center, NY, USA, and the Telefonica Research, Barcelona. He is currently an Assistant Professor with the Department of Computer Science, National University of Singapore. His current research interests include distributed systems and network economics. He was a co-recipient of the Best Paper Award in the IEEE Workshop on Smart Data Pricing 2015, the IEEE ICNP 2014, and the IEEE IC2E 2013.

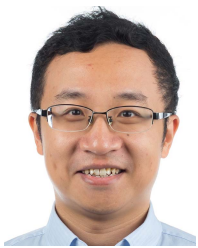

**Tom Z. J. Fu** received the B.Eng. degree in information engineering from Shanghai Jiao Tong University in 2006 and the M.Phil. and Ph.D. degrees from the Department of Information Engineering, The Chinese University of Hong Kong, in 2008 and 2013, respectively. In 2013, he joined the Advanced Digital Sciences Center, a Singapore-Based Research Center established by the University of Illinois at Urbana–Champaign, where he is currently a Senior Research Scientist and a Data Analytics Programme Manager. His research interests include cloud computing, real-time distributed stream analytics, software defined networking (SDN), Internet measurement and monitoring, peer-to-peer content distribution, and academic social network.